\documentclass[english]{article}
\usepackage{babel}
\usepackage{units}
\usepackage{amsmath}
\usepackage{esint}
\usepackage{mathptmx}
\usepackage{cite}

\usepackage [autostyle, english = american]{csquotes}
\MakeOuterQuote{"}

\begin{document}
	
	\title{CHSH Inequality on a single probability space}

	\author{C. M. Care \\
		Materials Research Institute, Sheffield Hallam University
		\\ Howard Street, Sheffield, S1 1WB UK \\
		c.m.care@shu.ac.uk }
	
	\maketitle

\begin{abstract}
Khrennikov and co-workers  have suggested in a series of papers that it is inappropriate to combine data from different experiments
when undertaking experimental tests of Bell's inequalities. They suggest that a correct analysis,
using a single probability space, leads to inequalities
which are not violated by experiment. If correct, this would be contrary to the normal interpretation of such experimental data.

However, in this note, a generalised Clauser-Horne-Shimony-Holt (CHSH) inequality
is derived for a system of four experiments constructed on a single probability space which combines the data from the four experiments.
In contradiction to Khrennikov \emph{et al}, it is shown that this leads to the standard CHSH inequality which is normally used to interpret experimental data. Thus the commonly accepted conclusion that experimental violations of Bell's
inequality imply that local realistic models are inconsistent with
the predictions of quantum mechanics has not been challenged.

\end{abstract}

\section{Introduction}

In a series of papers Khrennikov and co-workers \cite{Avis:2009.294,Khrennikov:2008.19,Khrennikov:2009.492,Khrennikov:2014.012019} have made strong claims that the experimental violation of the CHSH inequalities is a consequence of an erroneous use of experimental data.  It is suggested that the experimental violations of the CHSH inequality arise through incorrectly combining data from a number of experiments and suggest that mixing the data from different experimental contexts is an improper use of statistical data. 

In this paper we dispute these conclusions. We use the single probability space proposed by Khrennikov \emph{et al} and demonstrate that the central assumption they make about the CHSH inequality for quantum systems is too weak. The analysis presented below shows that a correct analysis of a model based on a single probability space yields the standard CHSH inequalities which are violated experimentally and this violation is \textbf{not} a consequence of incorrect use of the experimental data. 

In more recent papers Khrennikov  asserts that the error arises because of the confusion between conditional and absolute correlations (\emph{eg} page 772 of \cite{Khrennikov:2015.711} and page 7 of \cite{Khrennikov:2014.012019}). Once again the analysis presented below shows that the conditional correlations obey the standard CHSH inequality and the absolute correlations obey a stronger version of the CHSH inequality. Since there is a simple mathematical relationship between the conditional and absolute correlations and the inequalities which they obey, any circumstances which lead to one inequality being violated will also lead to the other being violated.

Avis \emph{et al} \cite{Avis:2009.294} give a detailed analysis of a EPR-Bohm-Bell experiments and
suggests a single probability space for the analysis of such
experiments; they describe this as a \emph{proper probability space}. It is also described as a \emph{common Kolmogorov probability
space} in Khrennikov \cite{Khrennikov:2014.012019}. In \cite{Khrennikov:2014.012019,Avis:2009.294} they assume a CHSH inequality of the form
\begin{equation}
\mid<A^{(1)},B^{(1)}>-<A^{(1)},B^{(2)}>+<A^{(2)},B^{(2)}>+<A^{(2)},B^{(1)}>\mid\quad\leq\quad2\label{eq:Avis form of CHSH inequality}
\end{equation}
 for random variables $A^{i}$ and $B^{i}$ defined on a single Kolmogorov probability space which they construct to describe four separate experiments; $A^{i}$ and $B^{i}$  and are defined below. They then claim that the "Bellian" variables $a_{i}$ and $b_{i}$ used to describe a single experiment obey the modified CHSH inequality (\emph{eg} equation (2) in \cite{Avis:2009.294})
\begin{equation}
\mid<a_{(1)},b_{(1)}>-<a_{(1)},b_{(2)}>+<a_{(2)},b_{(2)}>+<a_{(2)},b_{(1)}>\mid\quad\leq\quad 8\label{eq:Avis form of CHSH inequality Bellian}
\end{equation}
and that this inequality is not violated experimentally.

In this paper we show that the CHSH inequalities (\ref{eq:Avis form of CHSH inequality}) and (\ref{eq:Avis form of CHSH inequality Bellian}) are too weak and therefore lead to erroneous conclusions. We show that a correct analysis modifies the expression
(\ref{eq:Avis form of CHSH inequality}) to make it more restrictive, giving the modified form of the CHSH inequality shown in equations (\ref{eq:CHSH - final form}) or (\ref{eq:Generalised CHSH}), for the correlations of the random variables $A^{i}$ and $B^{i}$. It follows from this new expression that the standard CHSH inequality, equation (\ref{eq:Standard CHSH}), is recovered for the correlation functions  $<a_{i},b_{i}>$  which are normally considered in single experiments and this inequality \emph{is} violated experimentally. The correlation function $<a_{i},b_{i}>$ is equivalent to the conditional correlations defined by Khrennikov \cite{Khrennikov:2015.711,Khrennikov:2014.012019}.

\section{Background}

Bell \cite{Bell:1966.447}, and subsequently other workers, established
a number of inequalities which should be obeyed by local hidden variable
models of quantum mechanics; see \cite{Bell:1987} for a collection
of Bell's publications and also Clauser \emph{et al} \cite{Clauser:1969.880}. The inequality
derived in the latter paper will be referred to here as the CHSH inequality.
Bell type inequalities have been extensively tested experimentally
and it is found that the inequalities are normally violated by quantum
mechanical systems. Thus, although there remain some possible but
unlikely loopholes, there is now a consensus that local realistic
models are inconsistent with quantum mechanics. Two recent reviews of Bell's theorem and the associated experimental
work are to be found in \cite{Goldstein:2011.8378} and \cite{Brunner:2014.419}.

The canonical system to which Bell's result is typically applied is
a modification suggested by Bohm \cite{Bohm:1951} of the EPR experiment proposed
by Einstein \emph{et al} \cite{Einstein:1935.777} which measures the spin of two entangled
particles rather than their position; we refer to such experiments
as EPR-Bohm-Bell experiments.
\section{Hidden variables on a single probability space}

The work in this paper is based directly on the arguments Avis \emph{et al } 
and Khrennikov  used \cite{Avis:2009.294,Khrennikov:2014.012019} to construct a single Kolmogorov
probability space for the CHSH experiments. However, we come to different conclusions. We use the random variables set out in Section 3, \emph{Proper random experiment}, of \cite{Avis:2009.294}  
and derive results which show that the inequality (2) in \cite{Avis:2009.294} (\emph{ie} equation (\ref{eq:Avis form of CHSH inequality Bellian}) above ) is in error by being too weak. 

We consider an EPR-Bohm-Bell
experiment in which pairs of spin $\nicefrac{1/}{2}$ particles are
created and subsequently separate in opposite directions $A$ and
$B$. In each direction there are assumed to be two spin measurement
devices aligned in orientations $\mathbf{a}_{i};$$\;$i=1,2 and
$\mathbf{b}_{j};$$\;$i=1,2 on sides $A$ and $B$ respectively.
These devices measure the spin orientation of the particle in direction
$\mathbf{a}_{i}$ or $\mathbf{b}_{j}$. We assume that the
measurement devices are idealised and only yield outputs of $\pm1$
representing the possible outcomes of the measurements for a spin
$\nicefrac{1}{2}$ particle. Once the particles are separated
it is supposed that there is a device which randomly
determines which one measurement device will be used on side $A$,
$\mathbf{a}_{1}$ and $\mathbf{a}_{2}$. Similarly a device
randomly determines which device is used on side $B$, $\mathbf{b}_{1}$
or $\mathbf{b}_{2}$. It is supposed that the events of determining which measurement devices are used on each side are space-like separated.
Such a choice will limit the interaction between the experiments on sides $A$ and $B$.
With this arrangement there are therefore four possible experiments
which are undertaken depending on the random choice of ($\mathbf{a}_{1}$ or
$\mathbf{a}_{2}$) and ($\mathbf{b}_{1}$ or  $\mathbf{b}_{2}$).

The central problem is to understand the mechanism by which the space
like separated particles give correlated responses to the measurement
devices. At its simplest, is found that if $\mathbf{a}_{1}$ and
$\mathbf{b}_{1}$ have the same orientation they will always yield
anti-correlated results when those two devices are selected. If the devices have non-parallel orientation, the
response is more complicated but successfully described by quantum mechanics. We follow
Bell's argument to seek a local realistic explanation of this phenomenon.
We assume that at the time of creation of the pair of particles they
are each invested with information in the form of ``hidden variables''
which they carry to the detectors; these variables determine their
response to the detectors they encounter and must ensure the observed
correlations. This assumption provides a ``local'' mechanism by
which we might understand how the subsequent space-like separated
measurements are correlated in the absence of faster than light signalling.

It is reasonable to suppose that the hidden variables must be stochastic
since the observed responses to the measurement devices are stochastic
and the response of the detectors is determined by the hidden variables
which are created at the time of the pair creation. Indeed it would
be difficult to introduce a stochastic element once the particles
are sufficiently separated and still retain the observed correlations.
Following Bell \cite{Bell:1964.195} we will assume that there is a random
variable $\lambda$ which determines the outcomes of the possible
measurements. It is assumed that $\lambda$ is described by a probability
distribution $\rho(\lambda)$ normalised such that
\[
\intop\rho(\lambda)d\lambda=1
\]
although it is not necessary to specify the precise mathematical nature
of $\lambda$. We introduce random variables $\mu_{A}=1,2$ and $\mu_{B}=1,2$
to govern the process by which the detectors $\mathbf{a}_{i}$
and $\mathbf{b}_{i}$ are connected to the particle source. Thus, for example, when $\mu_{A}=1$ the detector $\mathbf{a}_{1}$ is connected to the source.
We assume a
probability measure $p_{\mu_{A},\mu_{B}}$
\[
\sum_{\mu_{A}=1}^{2}\sum_{\mu_{B}=1}^{2}\;p_{\mu_{A},\mu_{B}}=1
\]
 where, for example, $p_{1,2}$ is the probability that $\mu_{A}=1$
and $\mu_{B}=2$. Avis \emph{et al} \cite{Avis:2009.294} assume
\[
p_{\mu_{A},\mu_{B}}=\frac{1}{4}\qquad\forall\mu_{A},\mu_{B}
\]
and we will use this value initially in the following argument.

The random variable $\mu_{A}$ and $\mu_{B}$ determine which detectors
are used in any given experiment and $\lambda$ determines the outcome
of the measurement. Thus we have defined a two dimensional sample
space $\Omega=\{\lambda,(\mu_{A},\mu_{B}))\}$ which can be considered
to be an element of a single Kolmogorov probability space
\[
P=(\varOmega,\mathcal{F},\mathbf{P})
\]
where $\mathcal{F}$ is the $\sigma$ algebra associated with $\varOmega$
and $\mathbf{P}$ is the probability measure on $\mathcal{F}$. We
note that the hidden variable $\lambda$ is defined on a sample subspace
$\Omega_{0}=\{\lambda\}\in\Omega$.

In order to set the problem in its physical context we assume that
we are able to generate a large ensemble of pairs of particles. Each
particle pair is assumed to sample a value for $\lambda$ from the
distribution $\rho(\lambda)$ and this choice of $\lambda$ specifies
the outcome of the possible measurements which we represent by the
four random variables $a_{i}(\lambda); i=1,2$ and $b_{j}(\lambda); j=1,2$.
These are defined on the sample space $\Omega_0$ where, for example,
$a_{i}(\lambda)$ corresponds to the output for the detector
$\mathbf{a}_{i}$ for the given value of the hidden variable $\lambda$;
$a_{i}(\lambda)$ and $b_{j}(\lambda)$ are the Bellian variables
introduced by Avis. The variables $a_{i}(\lambda)$ and $b_{j}(\lambda)$ take only the values $\pm 1$
corresponding to the outputs of the detectors with orientation $\mathbf{a}_{i}$ or $\mathbf{b}_{j}$ for any given value of $\lambda$ . We have used the notation $\lambda$ to
be consistent with the notation used by Bell  rather than the notation
$\omega$ Avis used to represent an equivalent random variable.

In order to make contact with the definitions used in Section 3 of Avis \emph{et al} \cite{Avis:2009.294} we define four random
variables $A^{i}(\lambda,\mu_{A})$ and $B^{i}(\lambda,\mu_{B})$
by
\begin{eqnarray}
A^{i}(\lambda,\mu_{A}) & = & \delta_{\mu_{A},i}\:a_{i}(\lambda)\quad i=1,2 \nonumber \\
B^{j}(\lambda,\mu_{B}) & = & \delta_{\mu_{B},j}\:b_{j}(\lambda)\quad j=1,2 \label{eq:A B definitions}
\end{eqnarray}
where $\delta_{\mu,\nu}$ is the Kronecker delta function. Thus the
variables $A^{1}(\lambda,\mu_{A})$ has the following properties:-
\begin{enumerate}
\item $A^{1}(\lambda,\mu_{A})=\pm1$, if \emph{$\mu_{A}=1$}
\item $A^{1}(\lambda,\mu_{A})=0$, if \emph{$\mu_{A}=2$}.
\end{enumerate}
with similar definitions for $A^{2}(\lambda,\mu_{A})$ and $B^{j}(\lambda,\mu_{B})$
for $j=1,2$. The variables $A^{i}(\lambda,\mu_{A})$ and $B^{i}(\lambda,\mu_{B})$
are equivalent to the random variables $A^{(i)}(\omega)$ and $B^{(j)}(\omega)$
introduced by Avis.

\section{Derivation of the CHSH inequality }

We now explore the derivation of the CHSH inequality (\ref{eq:Avis form of CHSH inequality})
given above. The expression
makes use of the correlation function
\begin{equation}
<A^{(i)},B^{(j)}>=\sum_{\mu_{A}=1}^{2}\sum_{\mu_{B}=1}^{2}p_{\mu_{A},\mu_{B}}\intop A^{(i)}(\lambda,\mu_{A})B^{(j)}(\lambda,\mu_{B})\rho(\lambda)d\lambda\label{eq:A B Correlation function}
\end{equation}
where $\rho(\lambda)$ is the probability distribution for the hidden
variables $\lambda$ and we sum over the possible values of $\mu_{A}$
and of $\mu_{B}$; thus in equation (\ref{eq:A B Correlation function})
we have integrated and summed over the complete sample space $\Omega_{CMC}=\{\lambda,(\mu_{A},\mu_{B}))\}$.
We now attempt to recover the CHSH inequality by following the standard
arguments; see \emph{eg} page 37 of \cite{Bell:1987}.

\begin{multline*}
  <A^{(1)},B^{(1)}>-<A^{(1)},B^{(2)}>\nonumber = \\
  \sum_{\mu_{A}=1}^{2}\sum_{\mu_{B}=1}^{2}p_{\mu_{A},\mu_{B}}\intop[(A^{(1)}(\lambda,\mu_{A})B^{(1)}(\lambda,\mu_{B})\\
  -A^{(1)}(\lambda,\mu_{A})B^{(2)}(\lambda,\mu_{B})]\rho(\lambda)d\lambda\nonumber
\end{multline*}

In the analysis on page 37 of \cite{Bell:1987} the following modification of the RHS is made:-

\begin{multline}
 =\sum_{\mu_{A}=1}^{2}\sum_{\mu_{B}=1}^{2}p_{\mu_{A},\mu_{B}}\{\\
 \intop(A^{(1)}(\lambda,\mu_{A}),B^{(1)}(\lambda,\mu_{B})[1\pm(A^{(2)}(\lambda,\mu_{A}),B^{(2)}(\lambda,\mu_{B})]\rho(\lambda)d\lambda\\
  -\intop A^{(1)}(\lambda,\mu_{A}),B^{(2)}(\lambda,\mu_{B})[1\pm A^{(2)}(\lambda,\mu_{A}),B^{(1)}(\lambda,\mu_{B})]\rho(\lambda)d\lambda\}\label{eq:A B CHSH Derivation}\\
\end{multline}
However this introduces products of the form
\begin{equation*}
A^{1}(\lambda,\mu_{A}) A^{2}(\lambda,\mu_{A}) \quad \mbox{and}\quad B^{1}(\lambda,\mu_{B})  B^{2}(\lambda,\mu_{B})
\end{equation*}
which are always
zero and therefore the subsequent analysis in \cite{Bell:1987} is not valid. This arises because, for example, if $\mu_{A}=1$, $A^{2}(\lambda,\mu_{A})=0$ and if $\mu_{A}=2$,
$A^{1}(\lambda,\mu_{A})=0$; these represent the experimental arrangement
in which $A^{1}(\lambda,\mu_{A})$ and $A^{2}(\lambda,\mu_{A})$ correspond
to different experiments with different detectors $\boldsymbol{a}_{1}$
and $\boldsymbol{a}_{2}$ and these therefore never operate simultaneously.
Similar arguments apply to side $B$. Thus the derivation \cite{Bell:1987}
of the CHSH inequalities must be modified when considering the single probability space
and cannot be undertaken directly in terms of
the random variables $A^{i}(\lambda,\mu_{A})$ and $B^{j}(\lambda,\mu_{B})$.

In order to proceed, we use the definitions (\ref{eq:A B definitions}) to find
\begin{eqnarray}
 & < & A^{(i)},B^{(j)}>=\sum_{\mu_{A}}\sum_{\mu_{B}}p_{\mu_{A},\mu_{B}}\intop A^{(i)}(\lambda,\mu_{A})B^{(j)}(\lambda,\mu_{B})\rho(\lambda)d\lambda\nonumber \\
 & = & \sum_{\mu_{A}=1}^{2}\sum_{\mu_{B}=1}^{2}p_{\mu_{A},\mu_{B}}\;\delta_{\mu_{A},i}\;\delta_{\mu_{B},j}\;\intop a_{i}(\lambda)b_{j}(\lambda)\rho(\lambda)d\lambda\nonumber \\
 & = & p_{i,j}\intop a_{i}(\lambda)b_{j}(\lambda)\rho(\lambda)d\lambda\nonumber \\
 & = & p_{i,j}<a_{i},b_{j}>\label{eq:Correlation relation-1}
\end{eqnarray}
The quantity $<A^{(i)},B^{(j)}>$ is less than $<a_{i},b_{j}>$ because the variables $A^{(i)}$ and $B^{(j)}$ can be zero whereas $a_{i}$ and $b_{j}$ can only have values of $\pm 1$. 

In order to make contact with Khrennikov \emph{et al} \cite{Khrennikov:2014.012019,Khrennikov:2015.711}), we note that equation (\ref{eq:Correlation relation-1}) is consistent with the quantity $<A^{(i)},B^{(j)}>$ being identified as an \emph{unconditional} or \emph{absolute correlation}  and $<a_{i},b_{j}>$ being identified as a \emph{conditional correlation} where the condition is the choice of experiment $\{i,j\}$. This is in agreement with equation (4) in \cite{Avis:2009.294} and the result (\ref{eq:Correlation relation-1}) is also identical with the equation:-
\begin{equation}\label{eq:Khrennikov correlation}
C_{i,j}=p(a=i,b=j)Q_{i,j}
\end{equation}
given on page 721 of \cite{Khrennikov:2015.711}, where $C_{i,j}\equiv <A^{(i)},B^{(j)}>$ is termed a \emph{classical correlation} and $Q_{i,j} \equiv <a_{i},b_{j}> $ is termed a \emph{quantum correlation}. Further, if we assume $p_{\mu_{A},\mu_{B}}=\nicefrac{1}{4}$, equation (\ref{eq:Correlation relation-1}) becomes the result obtained in Section 5 of \cite{Avis:2009.294}.

We note that $a_{i}(\lambda)$ and $b_{j}(\lambda)$ are defined on the
subspace $\Omega_{0}$ of $\Omega$ and hence the correlation function
\[
<a_{i},b_{j}>=\intop a_{i}(\lambda)b_{j}(\lambda)\rho(\lambda)d\lambda
\]
 is well defined since they are each non-zero functions of $\lambda$.
We are now able to make the expansion

\begin{eqnarray}
 & < & A^{(1)},B^{(1)}>-<A^{(1)},B^{(2)}>\nonumber \\
 & = & \sum_{\mu_{A}}\sum_{\mu_{B}}p_{\mu_{A},\mu_{B}}\intop[(A^{(1)}(\lambda,\mu_{A})B^{(1)}(\lambda,\mu_{B})-A^{(1)}(\lambda,\mu_{A})B^{(2)}(\lambda,\mu_{B})]\rho(\lambda)d\lambda\label{eq:CHSH for A and B}\nonumber\\
 & = & \frac{1}{4}\intop[(a_{1}(\lambda)b_{1}(\lambda)-a_{1}(\lambda)b_{2}(\lambda)]\rho(\lambda)d\lambda
\end{eqnarray}

If we now follow the expansion in equation (\ref{eq:A B CHSH Derivation})
\begin{multline}
<A^{(1)},B^{(1)}>-<A^{(1)},B^{(2)}>   \\
 = \frac{1}{4}\intop(a_{1}(\lambda)b_{1}(\lambda)[1\pm(a_{2}(\lambda)b_{2}(\lambda)]\rho(\lambda)d\lambda\label{eq:Bellian correlation relation}\\
  - \frac{1}{4}\intop(a_{1}(\lambda)b_{2}(\lambda)[1\pm(a_{2}(\lambda)b_{1}(\lambda)]\rho(\lambda)d\lambda
\end{multline}
the products implicit in equation (\ref{eq:Bellian correlation relation})
are now well defined and non-zero. We may follow the standard argument
and take the absolute value of both sides and then apply the triangle
inequality to find:-

\begin{multline*}
\mid<A^{(1)},B^{(1)}>-<A^{(1)},B^{(2)}>| \quad \leq\\
  \frac{1}{4}\mid\intop(a_{1}(\lambda)b_{1}(\lambda)[1\pm(a_{2}(\lambda)b_{2}(\lambda)]\rho(\lambda)d\lambda\mid\\
  +  \frac{1}{4}\mid\intop(a_{1}(\lambda)b_{2}(\lambda)[1\pm(a_{2}(\lambda)b_{1}(\lambda)]\rho(\lambda)d\lambda\mid
\end{multline*}

We know that $[1\pm(a_{2}(\lambda)b_{2}(\lambda)]\rho(\lambda)$ and
$[1\pm(a_{2}(\lambda)b_{1}(\lambda)]\rho(\lambda)$ are both non-negative
and hence
\begin{multline*}
\mid<A^{(1)},B^{(1)}>-<A^{(1)},B^{(2)}>|  \leq\\
  \frac{1}{4}\intop\mid(a_{1}(\lambda)b_{1}(\lambda)\mid[1\pm(a_{2}(\lambda)b_{2}(\lambda)]\rho(\lambda)d\lambda\\
  +  \frac{1}{4}\intop\mid(a_{1}(\lambda)b_{2}(\lambda)\mid[1\pm(a_{2}(\lambda)b_{1}(\lambda)]\rho(\lambda)d\lambda
\end{multline*}

But we have that $\mid a_{i}(\lambda)\mid\leq1$ and $\mid b_{i}(\lambda)\mid\leq1$
so that the right hand side must be less than or equal to
\[
\frac{1}{4}\intop[1\pm(a_{2}(\lambda)b_{2}(\lambda)]\rho(\lambda)d\lambda+\frac{1}{4}\intop[1\pm(a_{2}(\lambda)b_{1}(\lambda)]\rho(\lambda)d\lambda
\]

and we may rewrite this as

\begin{multline*}
\frac{1}{2}\pm \frac{1}{4}[\intop(a_{2}(\lambda)b_{2}(\lambda)\rho(\lambda)d\lambda+\intop(a_{2}(\lambda)b_{1}(\lambda)]\rho(\lambda)d\lambda]\\
=\frac{1}{2}\pm\frac{1}{4}[<a_{2},b_{2}>+<a_{2},b_{1}>]\\
\end{multline*}

Hence we have
\begin{equation}
\mid<A^{(1)},B^{(1)}>-<A^{(1)},B^{(2)}>|\quad \leq \quad \frac{1}{2}\pm\frac{1}{4}[<a_{2},b_{2}>+<a_{2},b_{1}>]\label{eq:Penultimate result}
\end{equation}

We make use of the relation $<A^{(i)},B^{(j)}>=\nicefrac{1}{4}<(a_{i},b_{j}>$
between the Avis and Bellian variables from equation (\ref{eq:Correlation relation-1})
to write
\[
\mid<A^{(1)},B^{(1)}>-<A^{(1)},B^{(2)}>|\quad\leq\quad\frac{1}{2}\pm[<A^{(2)},B^{(2)}>+<A^{(2)},B^{(1)}>]
\]

or
\[
\mid<A^{(1)},B^{(1)}>-<A^{(1)},B^{(2)}>|\quad\leq\quad\frac{1}{2}-\mid[<A^{(2)},B^{(2)}>+<A^{(2)},B^{(1)}>]\mid
\]
rearranging and using the triangle inequality gives the result
\begin{equation}
\mid<A^{(1)},B^{(1)}>-<A^{(1)},B^{(2)}>+<A^{(2)},B^{(2)}>+<A^{(2)},B^{(1)}>\mid\quad\leq\quad\frac{1}{2}\label{eq:CHSH - final form}
\end{equation}
This is a stronger condition than equation (\ref{eq:Avis form of CHSH inequality}) cited by Avis and Khrennikov \cite{Avis:2009.294,Khrennikov:2014.012019}.
If we use the result $<A^{(i)},B^{(j)}>=\nicefrac{1}{4}<(a_{i},b_{j}>$
it also leads to a standard form of the CHSH inequality used to analyse experiments. \emph{ie}
\begin{equation}
\mid<a_{1},b_{1}>-<a_{1},b_{2}>+<a_{2},b_{2}>+<a_{2},b_{1}>\mid\quad\leq\quad2
\label{eq:Standard CHSH}
\end{equation}
This is in direct contradiction to the claims in \cite{Avis:2009.294,Khrennikov:2014.012019}. 

In a more recent paper, Khrennikov \cite{Khrennikov:2015.711} gives in his equation (33) a modified version of equation (\ref{eq:Avis form of CHSH inequality Bellian}) in which the factor 8 on the right hand side is replaced by 4. As has been shown above this result is also too weak; the factor should neither 8 nor 4 but 2 giving the standard CHSH result used in experimental analysis of EPR experiments. 

\section{Generalised CHSH inequality}

We may generalise the result (\ref{eq:CHSH - final form}) for the
case when $p_{\mu_{A},\mu_{B}}\neq\nicefrac{1}{4}$ by modifying equation
(\ref{eq:CHSH for A and B}) as follows:-
\begin{eqnarray*}
 &  & <\frac{1}{p_{1,1}}A^{(1)},B^{(1)}>-\frac{1}{p_{1,2}}<A^{(1)},B^{(2)}>\\
 & = & \sum_{\mu_{A}}\sum_{\mu_{B}}p_{\mu_{A},\mu_{B}}\intop[\frac{1}{p_{1,1}}(A^{(1)}(\lambda,\mu_{A})B^{(1)}(\lambda,\mu_{B}))\\
 & & -\frac{1}{p_{1,2}}(A^{(1)}(\lambda,\mu_{A})B^{(2)}(\lambda,\mu_{B}))]\rho(\lambda)d\lambda\\
 & = & \intop[(a_{1}(\lambda)b_{1}(\lambda)-a_{1}(\lambda)b_{2}(\lambda)]\rho(\lambda)d\lambda
\end{eqnarray*}

We may then follow the argument above in a straightforward way until
we find a modified version of equation (\ref{eq:Penultimate result})
\[
\mid\frac{1}{p_{1,1}}<A^{(1)},B^{(1)}>-\frac{1}{p_{1,2}}<A^{(1)},B^{(2)}>| \quad \leq \quad 2\pm[<a_{2},b_{2}>+<a_{2},b_{1}>]
\]

We now use the result (\ref{eq:Correlation relation-1}) to write

\begin{multline*}
\mid\frac{1}{p_{1,1}}<A^{(1)},B^{(1)}>-\frac{1}{p_{1,2}}<A^{(1)},B^{(2)}>| \quad \leq   \\
2 \pm [\frac{1}{p_{2,2}}<A^{(2)},B^{(2)}>+\frac{1}{p_{2,1}}<A^{(2)},B^{(1)}>] \\
\end{multline*}

 and hence equation (\ref{eq:CHSH - final form}) becomes
\begin{multline}
\mid \frac{1}{p_{1,1}}<A^{(1)},B^{(1)}>-\frac{1}{p_{1,2}}<A^{(1)},B^{(2)}> \\
+\frac{1}{p_{2,2}}<A^{(2)},B^{(2)}> +\frac{1}{p_{2,1}}<A^{(2)},B^{(1)}>\mid  \quad \leq \quad 2 \label{eq:Generalised CHSH} \\
\end{multline}
In the case that all the probabilities are equal to $\nicefrac{1}{4}$
this result becomes equation (\ref{eq:CHSH - final form}). As can be seen from equation (\ref{eq:Correlation relation-1}), the quantities $\nicefrac{<A_{i},B_{j}>}{p_{i,j}}$ are simply
the value of the expectation value $<(A^{(i)}B^{(j)}>$ conditional on the choice of the experiment $(i,j)$; this conditional probability is equal to  $<a_{i},b_{j}>$ and hence this result is identical to the standard CHSH result (\ref{eq:Standard CHSH}).

The result (\ref{eq:Generalised CHSH}) is a modified CHSH inequality applicable when four  separate experiments each with probability $p_{i,j}$ are analysed on a single probability space.

\section{Conclusions}

We can conclude therefore that the correct treatment of a single Kolmogorov
probability space for an EPR-Bohm-Bell experiment leads to a modified form of the CHSH inequality, equation (\ref{eq:Generalised CHSH}). This result is equivalent to the standard
CHSH inequality, equation (\ref{eq:Standard CHSH}) in contradiction to the claims in \cite{Avis:2009.294,Khrennikov:2008.19,Khrennikov:2009.492,Khrennikov:2014.012019,Khrennikov:2015.711}. 

The CHSH inequality (\ref{eq:Standard CHSH}) has been
shown experimentally to be violated (\emph{eg} \cite{Aspect:1982.91}) and this latter result is commonly interpreted to mean that it is not
possible to construct a local hidden variable model which is consistent
with quantum mechanics, or more importantly, experiment.

\end{document}